\documentclass[12pt,preprint]{aastex}

\begin{document}

\title{XMM Observations of an IXO in  NGC~2276}

\author{David S. Davis\altaffilmark{1}$^,$\altaffilmark{2}}

\and

\author{Richard F. Mushotzky\altaffilmark{3}}

\altaffiltext{1}{Joint Center for Astrophysics, Department of Physics,
University of Maryland,
Baltimore County, 1000 Hilltop Circle, Baltimore, MD 21250 }
\altaffiltext{2}{Laboratory for High Energy Astrophysics, Code 661,
Greenbelt, MD 20771}
\altaffiltext{3}{Laboratory for High Energy Astrophysics, Code 662,
Greenbelt, MD 20771}

\begin{abstract}
We present the results from a $\sim$53 ksec XMM observation of
NGC~2276.  This galaxy has an unusual optical morphology with the disk
of this spiral appearing to be truncated along the western edge. This
XMM observation shows that the X-ray source at the western edge is a
bright Intermediate X-ray Object (IXO). Its spectrum is well fit by a
multi-color disk blackbody model used to fit optically thick standard
accretion disks around black holes. The luminosity derived for this
IXO is 1.1$\times$10$^{41}$ erg s$^{-1}$ in the 0.5 - 10 keV band
making it one of the most luminous discovered to date.  The large
source luminosity implies a large mass black hole if the source is
radiating at the Eddington rate. On the other hand, the inner disk
temperature determined here is too high for such a massive object
given the standard accretion disk model. In addition to the IXO we
find that the nuclear source in this galaxy has dimmed by at least a
factor of several thousand in the eight years since the $ROSAT$ HRI
observations.

\end{abstract}


\section{Introduction}

A new class of X-ray emitting objects has recently been revealed in
nearby galaxies which are more luminous than normal X-ray binaries or
supernova remnants but do not attain the luminosities of conventional
AGN. These objects (Intermediate X-ray Objects: IXOs, Ptak 2002;
ultraluminous X-ray sources: ULXs, Makishima et al. 2000) are
generally found off-center of their host galaxy where supermassive
objects are not thought to reside.  The variability seen in these
objects indicate that they cannot be a collection of more normal
sources such as X-ray binaries or supernova remnants and therefore
must be a compact object.

The X-ray properties of IXOs distinguish them from the more common
point-like X-ray sources in spiral galaxies. They have luminosities
above that of most discrete compact sources in nearby galaxies. They
do not lie at the dynamic center (Long 1982; Fabbiano 1989; Marston et
al. 1995, Colbert \& Mushotzky 1999; Makishima et al. 2000) and have
luminosities well in excess of the Eddington luminosity from an accreting Neutron
star. These sources have luminosities in the range of 10$^{39}$ --
10$^{41}$ erg s$^{-1}$ (0.5 -- 10 keV), intermediate between X-ray
binaries accreting at the Eddington luminosity and most nuclear
sources.

While the basic emission mechanism is thought to be accretion, there is
an open question of whether these sources emit isotropically or if the
emission is beamed. If the emission is isotropic then the relation
between the X-ray luminosity and mass for a black hole accreting at
the Eddington rate is
$$L_{\rm B} = 1.5\times 10^{38}M/M_\sun\, {\rm erg\, s^{-1}}$$ when
the opacity is dominated by electron scattering in gas with solar
abundances. Interpreting this simply would imply that for the most
luminous IXOs the mass of the black hole must exceed 100 M$_\sun$.
The implication for very large mass black holes can be avoided if the
emission from the source is beamed in our direction (King et
al. 2001). In this case the mass of the central object can be reduced
so that normal accreting black holes or X-ray binaries can be the
source of the radiation. However, producing beamed emission from these
types of objects is still being investigated. For instance thick
accretion disks with a central funnel may radiate in excess of the
Eddington luminosity but a major theoretical question remains: is
beaming a natural consequence of high accretion rates in these
objects?

Here we present evidence for one of the more luminous IXOs known.
This object is located in the peculiar spiral galaxy NGC~2276 which is
undergoing intense star formation along the western edge of the
galaxy (Davis et al. 1997). This activity is thought to have been
triggered by a gravitational interaction with the elliptical galaxy
NGC~2300 (Gruendl et al. 1993; Davis et al. 1997). This object
has the properties of a ``classical'' IXO: it is located out in the
spiral arms of the galaxy away from the nucleus, has a super Eddington
luminosity, and exhibits long term time variability. Below we discuss
the XMM observation, the imaging and spectral analysis of the source.
Then we review the archival data for the source from the Einstein,
$ROSAT$, and $ASCA$ missions, and discuss the time
variability of the object.

\section{Observations}

NGC~2276 was observed for $\sim$53 ksec with XMM. The pointing center
for this observation was the center of the NGC~2300 group which is
approximately 7\arcmin$\,$ to the east of the elliptical NGC 2300.
We analyze only the MOS1 and
MOS2 data here since the source fell in a CCD gap on the PN camera.
Flares can be a significant problem with XMM data so we searched for
flares in the light curve for the full field and found only minor
flares near the end of the observation.  So for this analysis we use
the entire dataset.

The data were reduced and extracted using SAS 5.2, and the spectra are
analyzed using {\sc Xspec} 11.0.1. The latest calibration products were
used to determine the response files for the spectral analysis. The
extracted spectra are binned so that each channel has a minimum of
25 counts and the energy range is restricted to between 0.5 and 6.0
keV.

\section{X-ray Image}

The XMM observation clearly resolves the emission from NGC~2276 into
several components; the bright source to the west of the nucleus (the
IXO) at 7$^{\rm h}$ 26$^{\rm m}$ 49\fs 2 +85\arcdeg$\,$ 45\arcmin$\,$
55\farcs 2, two other point sources (XMMUJ 072718.8+854636 and XMMUJ
072816.7+854436) along with a more diffuse component from the galactic disk
of NGC~2276. 
The X-ray emission from the IXO may be slightly elongated to the southwest.
However, this elongation may be due to emission from the stellar complex
seen in the HST image but is much weaker than the IXO. Figure
1 shows a wider field of view which includes the sources XMMUJ 072718.8+854636
and XMMUJ 072816.7+854436.
Conspicuously absent is the nuclear source seen in the
$ROSAT$ HRI data (Davis et al. 1997). We have checked the astrometry
of the XMM field using the position of the nucleus of NGC 2300. We use
the NED position of NGC~2300 as the reference and find that we need to
shift the XMM MOS data to the north by 1\farcs 5. The XMM data in the
0.5 -- 10 keV band are shown as smoothed contours overlaying the DSS
image (Fig 1) and HST WFPC2 image of NGC~2276 (Fig 2). The brightest
source, the IXO, lies in the western side of the galaxy near the 
truncated edge.
In this observation we find no evidence for a point source near the
nucleus. For the combined dataset our upper limit for the nucleus
is $\sim$7.3$\times$10$^{-14}$ erg cm$^{-2}$ s$^{-1}$ or
$\sim$2.4$\times$10$^{39}$ erg s$^{-1}$ in the 0.5-10 keV band.
This relatively large upper limit is due to the intense  diffuse emission from
In the center of the galaxy.  We assume the distance to the group is 45.7
Mpc.

\section{Spectral Fits}

The spectrum of the IXO source has a total of 2344 source counts and
we fit the spectrum using an absorbed power law with the power
law index, normalization and foreground absorption as free parameters.
This fit results in an unacceptably high value of $\chi^2$ which is
due primarily to excess soft emission. Examination of figure 2 reveals
that there is diffuse emission associated with the disk of
NGC~2276. Instead of attempting to remove this spatially complex component
by modifying the
background spectra we instead model the disk emission as thermal
emission from a plasma and fit that component with the MekaL model in
{\sc XSPEC}, letting the temperature and intensity be free parameters 
and fixing the
abundance at the solar value. One consequence of this additional
spectral complexity is that we have to fix the total line-of-sight column
density to the Galactic value for the Milkyway 
(5.32$\times$ 10$^{20}$ atom cm$^{-2}$; Dickey \&
Lockman 1990) to obtain a stable fit for the temperature of the thermal
component.  Given the position of the X-ray source in NGC~2276, in
the outer disk and in the star forming regions of this morphologically
peculiar galaxy, we also explore models of emission from accreting
black holes.  Following Colbert \& Mushotzky (1999) we fit the data
with an absorbed power law and multi-color disk models (diskbb in
{\sc XSPEC}).  The power law model with the addition of the soft thermal
component is a substantial improvement over the simple absorbed power
law.  The multi-color disk model with the addition of the soft thermal
component is a better fit to the spectrum than a power law with reduced
$\chi^2$=1.05 (Figure 3) and no systematic residuals.
The luminosity of the MCD model for this source is
3.2$\times$10$^{40}$ erg s$^{-1}$ in the 0.5 -- 2.0 keV band,
which is unusually luminous for
black hole sources in nearby galaxies (Colbert \& Mushotzky 1999).

We also fit the source XMMUJ 072718.8+854636 with a power law and the 
multi-color
disk model discussed above. With 437 source counts either model fits
the data but the derived parameters for the power law model are 
rather unusual for
luminous galactic sources with
(N$_{\rm h}$ = 2.3$\times$10$^{21}$, $\Gamma$=3.5 and $\chi^2_\nu$ = 1.3 for
9 d.o.f.)
and thus we only discuss the MCD model here. The best fit MCD model 
has a reduced
$\chi^2$ of 0.8. The luminosity of the MCD model for this source is
4.2$\times$10$^{39}$ erg s$^{-1}$ extrapolated to the 0.5 -- 10.0 keV band (
3.8$\times$10$^{39}$ erg s$^{-1}$ also extrapolated to the 0.5 -- 2.0 keV band).
For the source XMMUJ 072816.7+854436, with 424 source counts,
we only fit a power law, which is often used to characterize the 
spectra of weak sources and using the best fit we find
that the luminosity in the 0.5 -- 10 keV band is 1.1$\times$10$^{39}$ 
erg s$^{-1}$
and 9.4$\times$10$^{38}$ erg s$^{-1}$ in the 0.5 -- 2.0 keV band assuming that
both sources are at the distance of NGC~2276.
Table 1 summarizes the fitting results.

\section{Source Variability}

We have extracted the light curve for the IXO from the XMM
MOS2 data. We selected these data because the MOS2 is the only detector
for which all of the emission from this source lies entirely on the chip.
With this restriction we do not have to allow for possible rate
variations due to different amounts of flux falling in the CCD gaps
during the long pointing.  We extract the count rate data for the
IXO and show the data grouped in
4600 second bins (figure 4). To search for periodicity in the
lightcurve we examined the power spectrum of the lightcurve with
frequency binning from 1 Hz to 3.3$\times$10$^{-4}$Hz.
Besides detecting the CCD readout time of
2.6 seconds (Ehle et al. 2001) no other significant period was detected.

To investigate the long term behavior of this object we have searched the
HEASARC database for archival observations and table 2 lists the observations
we use to investigate the long term trends for this object.
The Einstein IPC observed the NGC~2300 group for only 1940 seconds and this
exposure time was insufficient to detect the spiral galaxy. We determine
an upper limit of 2.0 $\times$10$^{40}$ erg s$^{-1}$ in the 0.5 -- 2.0 keV
band during that time.

The {\sl ROSAT} PSPC data, with its poorer resolution than the {\sl ROSAT} HRI
or the XMM data, potentially suffers from confusion with
other sources in the galaxy. However, the location of the peak of the
emission is distinct from the nucleus and  consistent with the XMM and
HRI position of the IXO.
During the $ROSAT$ PSPC observations the luminosity of this source in
the 0.5 -- 2.0 keV band was 2.5$\times$10$^{40}$ erg s$^{-1}$ in the
initial observation in 1992 and 2.2$\times$10$^{40}$ erg s$^{-1}$ in
the first of the two observations in August 1993, an interval of 1.3 
years.  This
is consistent with no change between April 25 1992 and Aug 22 1993.
The count rate in the PSPC data for the second 1993
observation shows that the source luminosity increased by a factor of roughly
2 between observations separated by only $\sim$3.25 hours to
4.4$\times$10$^{40}$ erg s$^{-1}$. These luminosities are determined
using the best fit XMM MCD model fit to the PSPC spectra with only the
normalizations allowed to vary. The hard and soft band data show that
the hardness ratio did not change appreciably between these
observations.

The $ROSAT$ HRI data originally analyzed by Davis et al. (1997)
were a combination of two HRI observations separated by about six months.
In the first observation the luminosity of the IXO is
3.7$\times$10$^{40}$ erg s$^{-1}$. By the second observation it had
decreased in luminosity to 2.4$\times$10$^{40}$ erg s$^{-1}$ in the
0.5 -- 2.0 keV band.
Even given the
assumptions that are needed to convert the HRI count rate to flux,
e.g. similar disk temperature at that epoch, this is clear evidence
that the source had dimmed significantly since the 1993 PSPC observations.

We have also examined the $ASCA$ GIS data for signs of variability from
NGC~2276 and
find variability of about a factor of 4 between the different
observations.  Unfortunately, with the GIS spatial resolution
($\approx 0.9\arcmin\,$ at 2 keV with large wings) we
cannot isolate the IXO so the luminosities are totals for
the galaxy. As we point out below the nuclear region in the galaxy is
also variable, so that the variability during this time may be the IXO, the
nucleus, or some combination of the two.  The best fit MCD model to
the XMM data yields a luminosity of 3.2$\times$10$^{40}$ erg s$^{-1}$
in the same band for the GIS data.

The archival data clearly show that this source is variable and that
while the flux may remain fairly constant for long periods of time
this IXO can change luminosity by a factor of 2 in a short time and
factors of 2 or 3 over a year or more. Table 3 lists the derived
luminosities and gives the 1$\sigma$ errors for the luminosity and
Figure 5 summarizes the source variability.
In addition to the IXO the nuclear source has also varied. The
$ROSAT$ HRI data show that the nuclear region had an X-ray luminosity of
$\sim$2.0$\times$10$^{40}$ erg s$^{-1}$ in the 0.5 - 2.0 keV band for
both observations. The XMM upper limit of the flux from the nuclear region is
$<$1.4$\times$10$^{39}$ erg s$^{-1}$ in the same band.

\section{Discussion}

The X-ray luminosity, variability, spectral fits, and position of the 
source in the
galaxy indicate that this is another example of an IXO.
Interpreting this source as a young X-ray bright supernova remnant or
a remnant evolving in a dense medium (Franco et al. 1993) is
problematic.  The HST image (figure 2) does not show that the source
is clearly associated with a diffuse source.
The luminosity of this source is 1.1$\times$10$^{41}$
erg s$^{-1}$ extrapolated to the 0.5 -- 10 keV band ( 3.2$\times$10$^{40}$ erg
s$^{-1}$ in the 0.5 - 2.0 keV band) which is consistent with the luminosity
of the source in March 1994 in the $ROSAT$ HRI data (3.7
$\times$10$^{40}$ erg s$^{-1}$ in the 0.5 - 2.0 keV band). Yet the X-ray
luminosity of supernova remnants are expected to evolve strongly over
such a timescale (Chevalier 1982).  The PSPC and $ASCA$ data
indicate that the luminosity of the source is not simply declining but
shows variability on time scales of both hours and years consistent
with this being a compact source rather than a SNR.

The unabsorbed luminosity of this source (1.1$\times$10$^{41}$ erg s$^{-1}$ in
the 0.5 -- 10 keV band) makes this one of the most luminous IXOs known to date.
Typical luminosities for these sources are 10$^{39}$ -- 10$^{40}$ erg s$^{-1}$
in the 0.5 -- 10 keV band while the most luminous of these sources seems to
be the one in M82 with a maximum luminosity of $\sim$10$^{41}$ erg s$^{-1}$
(Matsumoto et al. 2001).
However these objects are strongly variable with the X-ray luminosity
changing by an order of magnitude or more over a timescale of a few
years.  In fact the source in NGC~2276 has varied by a factor of 
$\sim$2 between
the $ROSAT$ HRI observation and the XMM observation. We have also
searched for short term variability in the XMM data by constructing a
power density spectrum (PDS). We binned the data with a bin width of
300 s, 500 s and 3000 s. Using the powspec tool from the FTOOLS
version 5.1 we found no significant structure in the PDS for the source
over the timescale of the XMM observation.

\subsection{Spectral Results}

The spectrum of this source is well fit by a multi-color disk model
(Mitsuda et al. 1984).  This is typical for such sources (Colbert \&
Mushotzky 1999; Makishima et al. 2000; Wang 2002) although a power law
spectrum seems to be more appropriate in some cases (Strickland et
al. 2001).  The multi-color disk model (Mitsuda et al. 1984) assumes a
standard optically thick accretion disk with the inner disk edge
located at the last stable orbit for a Schwarzchild black hole. The
spectrum is a superposition of multiple blackbody spectra up to a
maximum temperature (T$_{\rm in}$). This maximum temperature is
expected to occur at the inner edge (R$_{\rm in}$) of the accretion
disk. Under these assumptions the temperature at R$_{\rm in}$ can be
used to infer the radius of the last stable orbit and thus the mass of
the black hole.  The mass of the black hole can be found using the
measured temperature for the MCD model of 2.05 keV and the definition
of the {\sc Xspec} diskbb model normalization $K = ((R_{\rm
in}/km)/(D/10kpc))^2 * cos(\theta)$, where D is the distance to the
source and $\theta$ is the inclination angle of the accretion disk.
Solving for R$_{\rm in}$ and using eq 8 of Makishima et al. (2000) we
can estimate the mass of the black hole as $M = 516 (K/\alpha^2
cos^2(\theta))^{1/2} M_\sun$, where K is the normalization of the {\sc
Xspec} MCD model, $\theta$ is the inclination of the accretion disk,
and $\alpha$ is the ratio of the inner disk radius to that of the last
stable orbit for a Schwarzchild black hole, given that the {\sc Xspec}
normalization is 1.26$\times$10$^{-3}$ and assuming that $\alpha$=1, a
Schwarzchild black hole. If we assume that the inclination angle is
zero, a face-on system, then from the spectral data we can estimate
that the mass of the black hole is 18.3 M$_\sun$.  This is in contrast
to the mass derived via the Eddington limit of $\sim$730M$_\sun$. The
problem, as discussed by Makishima et al. (2000), is that the spectral
fits require a high inner disk temperature and a small black hole mass
while the requirement that the Eddington limit not be exceeded
requires a much larger black hole mass.  

One method of increasing the inner disk temperature and retaining a
larger mass black hole is to have a Kerr (spinning) black hole.  This
reduces the radius of the last stable orbit to at most half that of a
Schwarzschild black hole for prograde orbits. This allows a higher
inner disk temperature.  Simulations by Zhang, Chui \& Chen (1997)
show that for a maximally rotating Kerr black hole the inner disk
temperature can be a factor of $\sim$3 higher than in a Schwarzschild
black hole with the same mass and accretion rate which can bring the
spectrally measured temperature in line with the
mass derived from the Eddington limit. Modeling the effects of the
strong gravitational field near the black hole shows that for a
Schwarzschild singularity the cosine law is approximately valid but
for a Kerr black hole cos($\theta$) is restricted to about 0.17 -- 0.4
(Zhang et al. 2001). So, for a rapidly rotating Kerr black hole the
estimated mass range is (30 -- 40 $M_\sun)\alpha^{-2}$. 
The requirement that we match both the temperature of the
inner disk and the luminosity of the source requires that $\alpha \la$
0.26. This is consistent with the conclusions of Makishima et
al. (2000) that IXO's may be the result of accretion onto Kerr Black
Holes.

While is seems fairly certain that a black hole is the central engine
for these sources the exact mechanism that produces the unusual
luminosity in IXO's is still uncertain. The proposed models fall into
two main categories: beamed models and unbeamed models.  For the
beamed models the prediction is that this is a short-lived
evolutionary phase and should be associated with young stellar
populations (Taniguchi et al. 2000; King et al. 2001).  The position
of this IXO in the star forming regions along the truncated edge of
NGC~2276 is consistent with that interpretation.

\subsection{Beamed Emission Models}

One of the possible explanations for the incredible luminosity of the
IXOs is that the radiation is geometrically beamed. Essentially this
means that the true solid angle of the radiation is less than 4$\pi$
sr.  One method of achieving this is to have a thick disk partially
block the radiation and redirect a fraction of the initially isotropic
emission into a smaller solid angle. In this scenario some of the
X-ray radiation is absorbed by the thick disk and re-radiated in other
wavelength bands with some scattered into our line of sight. We stress
that no physically reasonable scattering region will be a perfect
mirror and thus a substantial fraction of the intrinsic luminosity
will be absorbed and re-radiated.  The most likely spectral regions to
observe this re-radiated energy would be either the optical or
the infrared band. Thus in naive geometrical beaming models one expects a
luminous optical-IR counterpart to the IXO.

We estimate the amount geometrical beaming using the thick accretion
disk model of Madau (1988). While this model is not unique it does
allow one to calculate the expected luminosity in other wavelength
bands and is the only one that we are aware of that calculates the
opening angle expected in a thick accretion disk.  Using figure 9 from
Madau (1988) we see that the beamed radiation is most intense within a
half opening angle of $\sim$20$\arcdeg$ of the rotation axis of the
torus. It predicts that an observer inside that half angle infers that
the source is radiating at roughly 16 times the "true" rate.  Assuming
that we are observing the source down the rotation axis of the funnel
then the implied observed flux is 16 times that of the true
isotropically radiated emission. That is, the observed flux would be
enhanced by a factor of 16 compared to the true flux, which is simply
the ratio of the opening angle to the total solid angle assuming two
cones. Thus, using this model the true isotropic luminosity of the IXO
in NGC2276 is 6$\times$10$^{39}$ erg s$^{-1}$ consistent with a black
hole mass of 45 M$_\sun$.

In the Madau (1988) model, part of the radiation is scattered and
enhances the surface brightness of the funnel walls and part of the
radiation is absorbed by the funnel walls; only a small fraction of
the intrinsically radiated luminosity is being captured by the
scattering cone and most of it is being radiated into the blocking
torus.  The difference between the isotropic luminosity predicted by
the Madau model and that from the geometric argument is the ratio of
the luminosity that goes up the funnel to that which is radiated in
4$\pi$ sterradians. A factor of 16 for a double cone model. So,
5$\times$10$^{39}$ erg s$^{-1}$ must be absorbed by the thick
accretion disk. Since this X-ray radiation is absorbed by the material
in the torus and it must be re-radiated in some other bandpass. The
most likely spectral region would be the optical or far-IR. This
would produce a luminosity that would result in an absolute magnitude
of $\approx$-11 without any color corrections, much brighter than the
most luminous stellar objects in any galaxy (excluding supernova)
which have absolute magnitudes in the range of -9 to -11 (Humphreys \&
Davidson 1979; Massey et al. 2000. These extremely luminous stars are
similar to Eta Carinae and would be easily identifiable as the
counterparts of the IXOs in nearby galaxies.  This is driven home by
studies of the brightest IXO, the source in M33 (Long, Charles \&
Dubus 2002), where there is no optical evidence, at the absolute
visual magnitude of $\sim$ -5.2, for an unusual counterpart to the
L$_{\rm x}\approx$10$^{39}$ erg s$^{-1}$ source. The lack of any such
identifications suggests to us that either the absorbed energy is
radiated in the far-IR, for which similar surveys have not been done,
or that geometrical collimation is not the process responsible for the
apparently high luminosities. While the above limits are model dependent
the lack of extremely luminous optical counterparts to the IXOs
strongly limits any geometrical collimating scenario. It is beyond the
scope of this paper to calculate the general restrictions,
e.g. albedo, solid angle of scattered, absorption models and fraction
of the luminosity that is intrinsically isotropic but we simply wish
to point out that the absence of an extremely luminous optical
counterpart strongly limits geometrical beaming models.

The second alternative is true beaming in which the intrinsic radiation
field is highly anisotropic. This has been observed in many
extragalactic radio sources (e.g. Bl Lac objects) and in several
Galactic objects (e.g. SS433 and the X-ray microquasars).  In
relativistic beaming models (e.g., Urry \& Shafer 1984) this is a
natural consequence of relativistic motion. The existence of a highly
anisotropic radiation field in non-relativistic objects is mainly the
subject of theoretical work and, so far, has not received observational
support from the extensive work on Seyfert galaxies and quasars. A
natural consequence of relativistic beaming is a connection between
the beaming factor $\Gamma$ and the solid angle of the beam. The
luminosity is enhanced by roughly $\Gamma^4$ and the solid angle of
the beam is of order $\Gamma^{-2}$. The number of sources is a strong
function of the intrinsic beaming and the intrinsic luminosity
function. In reviewing the cases discussed in Urry and Shafer (
1984), we were immediately struck by the fact that the beamed 
luminosity function is
the same as the unbeamed luminosity function; however, the sources are
more luminous by $\Gamma^4$.  Using this fact and the intrinsic
luminosity function of X-ray sources in external spiral galaxies having a
luminosity function with a slope of $\sim -$1.5 (e.g. Tennet et
al. 2001) we can predict the number of lower luminosity sources
from both the beamed and unbeamed populations. Of course at lower
luminosities there must be a substantial population of non-black hole
sources (e.g. high and low mass X-ray binaries). Given that we have
only one IXO in NGC~2276 (or at most three if the other two sources are
indeed IXO's) one would predict $\sim$30 times more sources
from a beamed component at a luminosity which is one tenth the
luminosity of the IXO. This is completely ruled out by the data. In
addition to the relativistically beamed sources one must observe the
parent population of these sources, those objects whose radiation is
not beamed in our direction. While we do not know if there is any
unbeamed radiation at all, if Bl Lacs are a guide, there should be
many such objects. The intrinsic numbers are roughly $\Gamma^2$ larger
than the beamed sources. So, if we have beaming factors of $\Gamma\sim
2.5$ and thus apparent luminosity enhancements of $\sim$80, reducing the
true luminosity to $\sim$10$^{39}$ and as such easily accommodated by a
20M$_\sun$ black hole, then we expect $\sim$10 times more unbeamed
sources of intrinsic luminosity $\sim$10$^{39}$.  These objects are
not seen either.  The detailed implications of the beaming scenario
are discussed in K$\ddot{o}$rding, Falcke \& Markoff (2002) and are in
good agreement with the qualitative discussion above. In their figure
1 they show the expected factor of 30 increase in the number density
of log L$_{\rm x}\sim$39.5 objects over those at log L$_{\rm
x}\sim$40.5 for the beaming models.  This large increase in the number
density in NGC~2276 is ruled out by our data.  Thus, the absence of
lower luminosity, ultra-luminous sources strongly constrains the
relativistic beaming models for this object.

It is also well known that relativistically beamed objects show very
large amplitude variability (e.g. the microquasars and Bl lac
objects).  In addition, the other sources that are known to be beamed
(e.g. BL Lac objects) show large amplitude variability on short
timescales and have PDS that can be represented by a power law,
$f^{-\alpha}$, with an index $\alpha$ of 2 to 3.  (e.g. Kataoka et al.
2001). This steep slope in the PDS can be interpreted as variability
of the source on short timescales.  The PDS for the IXO in NGC~2276 is
flat between $\sim$10$^{-5}$ and 5$\times$10$^{-3}$ Hz which is quite
different from the PDS for Bl Lac objects. If Galactic microquasars
are an analog of IXOs then the lack of short term variability may pose
a problem.  Multi-wavelength monitoring of GRS 1915+105, a Galactic
microquasar, shows that this source is very variable in the X-ray
band. GRS 1915+105 also demonstrates quasi-periodic flaring on a
timescale of about 2700 seconds and also flare activity during periods
of more constant luminosity (Ueda et al. 2002).  So, if the IXO in
NGC~2276 is a microquasar viewed along the jet then the lack of short
term variability is puzzling.  The very low amplitude of short
timescale variability in G1915+105 is also seen in the IXO's observed
by ASCA (Ueda et al. 2002) and is in strong contradiction to
expectations for relativistically beamed sources.

In addition to the temporal properties of the jets we also look at the
spectral properties. As K$\ddot{\rm o}$rding et al. (2001) point out the
jet hypothesis requires either rather powerful jets with reasonable Lorentz
factors or rather large Lorentz factors with less powerful jets. This
implies a rather unique spectrum for the jet dominated emission. The
only known jet X-ray sources, Bl Lac objects, are fit by power law
models with curvature at both low and high energies, a spectral shape
very different from the MCD models which fit the IXOs including the
source in NGC~2276. Thus models which can produce the X-ray spectra
from a jet in the IXOs (Georganopoulos, Aharonian \& Kirk 2002) need
to be rather different from those that work in Bl Lacs.

If IXOs are radiating isotropically at the Eddington limit then even
for the systems with more typical luminosity the mass of the
accreting object must exceed several hundred M$_\sun$. Previous
observations have found evidence for stellar mass black holes with
masses $\la$10 M$_\sun$, for supermassive black holes (M$\ga$10$^6$
M$_\sun$) and evidence for an intermediate mass black hole in at
least one stellar cluster (Gebhardt, Rich and Ho 2002).  These IXOs may
represent the first evidence that intermediate mass black holes are
more wide spread as suggested by Colbert \& Mushotzky (1999). Assuming
the existence of intermediate mass black holes the unbeamed models
require a black hole mass of $\geq$100 M$_\sun$ in a binary system
(King et al. 2001). One problem with this is that the inferred
temperature of the inner disk is inconsistent with such a high mass
black hole.  The high inner disk temperature can be explained either
by having the central source be a Kerr black hole (Makishima et
al. 2000) or by assuming non-standard accretion disk model
(e.g. Watarai et al. 2001).

We are thus left either with the idea of King \& Puchnarewicz (2002) that a
new mechanism must be at work which produces intrinsically beamed
non-relativisitic radiation fields or that these objects are truly
very massive black holes. If these objects represent a new type of
accretion phenomena one might expect other unusual spectral or temporal
properties such as are seen in the narrow line Seyfert galaxies
which have very distinct X-ray spectral and temporal properties from
normal Seyfert galaxies (Leighly 1999). One strong test of the idea
that these objects are very massive black holes will come from
measurement of the PDS of time variability.
Recent XTE results on both galactic and extragalactic black holes
(Nandra et al. 2000; Uttley, McHardy, Papadakis 2002) show that there is a
scaling between the knee of the PDS and the mass of
the black hole. If the IXOs are truly $\sim$100-1000 M$_\sun$ black holes then
their PDS will be intermediate between those of the 10 M$_\sun$ galactic
objects and the 10$^6$-10$^8$ M$_\sun$ AGN. This is easily testable with XMM
observations of the brightest IXOs.

This set of unusual properties for the X-ray binary population in
NGC2276 argues, as suggested by King et al (2001), that individual IXOs
may represent rare occurrences of very massive isolated black
holes. We note that the situation in the Antenna of several
ultra-luminous sources in a individual galaxy is quite rare (Colbert
and Ptak 2002) with an expectation value on the basis of XMM and Rosat
data of $\la$2 per galaxy and thus the situation in NGC~2276 may be
much more common (Foschini et al. 2002). Models which are driven
by the high rate of IXOs in some starburst galaxies may not be
applicable in general.

\subsection {The Central Source}

The nucleus of NGC~2276 shows no signs of AGN activity based on
the optical colors (Davis et al 1997) or spectra (Kennicutt
1992). However, the strong X-ray variability of the nuclear region of
NGC~2276 may be an example of large amplitude variability seen in a
few other AGN.  While detailed variability studies on times scales of
years are only available for the brightest 25 AGNs, it is not unusual
for objects to vary by factors of five over time scales of years and
factors of ten are not unknown (e.g., Peterson et al. (2000) for
NGC4051 and Weaver et al. (1996) for NGC2992). Thus it is not clear if
the pattern of variability of the central source in NGC2276 is
unusual. However, it is clear that the strong variability confirms that
this object is indeed an AGN. In the Rosat all sky survey there were
five galaxies that were in outburst, three of which show no signs of
nuclear activity in the optical band. Donely et al (2002) conclude
that the rate of large-amplitude X-ray outbursts from inactive
galaxies in the local Universe is $\sim$ 9.1 $\times$ 10$^{-6}$
galaxy$^{-1}$ yr$^{-1}$.  This rate is consistent with the predicted
rate of stellar tidal disruption events in such galaxies. It is
entirely possible that the NGC2276 nucleus is another example of this
sort of activity.

\section{Conclusions}

A long $\sim$53 ksec observation of the spiral galaxy NGC~2276 has
revealed the presence of a bright IXO at the western edge of the
galaxy. The spectrum of this source is fit best with a multi-color disk
blackbody with a thermal MekaL model included to account for the X-ray
emission from the disk of the galaxy.  The inferred temperature of the
inner disk is kT = 2.05 keV. This along with the high bolometric luminosity of
1.1$\times$10$^{41}$ erg s$^{-1}$ implies that either the singularity
must be rotating or that a non-standard disk model must be
employed. In addition, the strong star formation rate seen in this
disturbed galaxy lends weight to the proposal by Taniguchi et
al. (2000) that intermediate mass black holes can form in regions of
intense star formation.

IXO models where the radiation is geometrically beamed also predict
that in addition to the direct and reflected radiation, that part of
that radiation incident on the thick disk must be absorbed. This
absorbed radiation must be re-radiated in some energy band. Assuming
that the model of Madau (1988) is applicable we find that
$\sim$2.4$\times$10$^{40}$ erg s$^{-1}$ must be re-radiated. A source
with this level of emission would easily be seen as a extremely
luminous optical counterpart to in IXO's associated with nearby
galaxies.  So either this radiation must be re-radiated at far-IR
wavelengths or geometrical beaming is not at work in IXOs.

The models where IXOs are relatively normal X-ray binary systems that
are relativistically beamed in our direction seem to predict that
short term variability would be a characteristic of such systems.  We
searched for short term variability in this source and found none.
The power density spectrum for this IXO is flat and shows none of the
power law structure seen in other beamed sources. However, this IXO
does show time variability on timescales of hours and years. On the
timescale of years the variability can be up to factors of two or
three and it can vary by 20\% in only a few hours.  Thus, models where
IXO's are relatively normal X-ray binary systems that are beamed in
our direction seem less likely in this case.

A more stringent test to distinguish between various models of IXOs
requires that accurate positions of the object be known in order for
counterparts at other wavelengths to be found. This will also allow
for detailed studies of the environment in which these objects are found
and, presumably, formed. For this a more accurate X-ray position
from $Chandra$ is needed. The combination of the high throughput of
XMM combined with the positional accuracy of $Chandra$ is vital for
the further study of this object.

\acknowledgments

This research has made use of the NASA/IPAC Extragalactic Database
(NED) which is operated by the Jet Propulsion Laboratory, California
Institute of Technology, under contract with the National Aeronautics
and Space Administration. This research has also made use of data
obtained from the High Energy Astrophysics Science Archive Research
Center (HEASARC), provided by NASA's Goddard Space Flight Center. We 
would also like to thank the referee whose comments significantly 
improved the paper. 

\clearpage

\begin{deluxetable}{llc}
\footnotesize
\tablecaption{Model fits}
\tablewidth{0pt}
\tablehead{\colhead{Model} &\colhead{Parameters} & \colhead{$\chi^2$ / dof}}
\startdata
\multispan{3}{\hfil IXO\hfil}\\
Power Law& $\Gamma$=1.30$^{+0.09}_{-0.10}$    & 110/79 \\
          & N$_{\rm H}$=5.38$^{+2.22}_{-3.10}$ & \\
          & & \\
Power Law& $\Gamma$=1.40$^{+0.12}_{-0.17}$ &  85 / 76 \\
+        & kT=0.31$^{+0.10}_{-0.04}$ & \\
MekaL    & N$_{\rm H}$=17.8$^{+14.8}_{-12.8}\,$ & \\
          & & \\
Diskbb   & T$_{\rm in}=$2.05$^{+0.27}_{-0.27}$ & 80 / 76\\
          & K$_{\rm MCD} = 1.26\times 10^{-3}$ & \\
+        & N$_{\rm H}$=0.13$^{+9.59}_{-0.13}\,$  & \\
MekaL    & kT=0.35$^{+0.20}_{-0.06}$ & \\
\multispan{3}{}\\
\tableline
  & & \\
\multispan{3}{\hfil XMMUJ 072718.8+854636\hfil}\\
Diskbb   & T$_{\rm in}=$0.45$^{+0.20}_{-0.75}$ & 4.12 / 6\\
          & K$_{\rm MCD} = 3.96\times 10^{-4}$ & \\
+        & N$_{\rm H}$=$<$47.3$\,$  & \\
MekaL    & kT=0.35$^{+0.20}_{-0.06}$ & \\
\multispan{3}{}\\
\tableline
  & & \\
\multispan{3}{\hfil XMMUJ 072816.7+854436\hfil}\\
          & N$_{\rm H}$=5.32 & \\
Power Law& $\Gamma$=1.40$^{+0.67}_{-0.53}$ &  17.4 / 16 \\

\enddata
\tablenotetext{a}{Intrinsic N$_{\rm H}$ in units of $\times$10$^{20}$ 
atoms cm$^{-2}$}
\end{deluxetable}

\begin{deluxetable}{llccl}
\footnotesize
\tablecaption{Observation Log}
\tablewidth{0pt}
\tablehead{\colhead{Mission} &\colhead{Sequence \#}&
\colhead{Observation Date}&\colhead{MJD} & \colhead{Exposure (s)}}
\startdata
Einstein &I6645 &1980-03-04 &44302&1940 \\
$ROSAT$  &RP900161N00 &1992-04-25&48737& 5997 \\
$ROSAT$  &RP900512N00 &1993-08-22&49221& 9029\\
$ROSAT$  &RP900513N00 &1993-08-22&49221& 8417\\
$ROSAT$  &RH600498N00 &1994-03-18&49429& 52235\\
$ROSAT$  &RH600498A01 &1994-08-29&49593& 21731\\
$ASCA$   &80012000 &1993-05-29&49136& 19808\\
$ASCA$   &80013000 &1993-05-30&49137& 21472\\
$ASCA$   &85005000 &1997-10-31&50752& 44128\\
$ASCA$   &85005010 &1997-10-31&50752& 37296\\
$XMM$    &0022340201&2001-03-16&51984&53830 \\

\enddata
\end{deluxetable}
\begin{deluxetable}{lcc}
\footnotesize
\tablecaption{X-ray Luminosity}
\tablewidth{0pt}
\tablehead{\colhead{Mission} &\colhead{MJD}&
\colhead{L$_{\rm x}$\tablenotemark{a} }\\
\colhead{} & \colhead{} & \colhead{(10$^{40}$erg s$^{-1}$)} }
\startdata
Einstein &44302& 2.0 $\pm$ 0.6 \\
$ROSAT$  &48737& 2.5 $\pm$ 0.5 \\
$ROSAT$  &49221& 2.2 $\pm$ 0.5 \\
$ROSAT$  &49221& 4.4 $\pm$ 0.6 \\
$ROSAT$  &49429& 3.7 $\pm$ 0.6 \\
$ROSAT$  &49593& 2.4 $\pm$ 0.4 \\
$ASCA$   &49136& 4.3 $\pm$ 0.4 \\
$ASCA$   &49137& 1.8 $\pm$ 0.2 \\
$ASCA$   &50752& 5.3 $\pm$ 0.5 \\
$ASCA$   &50752& 4.4 $\pm$ 0.5 \\
$XMM$    &51984& 3.2 $\pm$ 0.3 \\
\enddata
\tablenotetext{a}{The errors given are 1$\sigma$}
\end{deluxetable}

\clearpage

\begin{figure}
\plotone{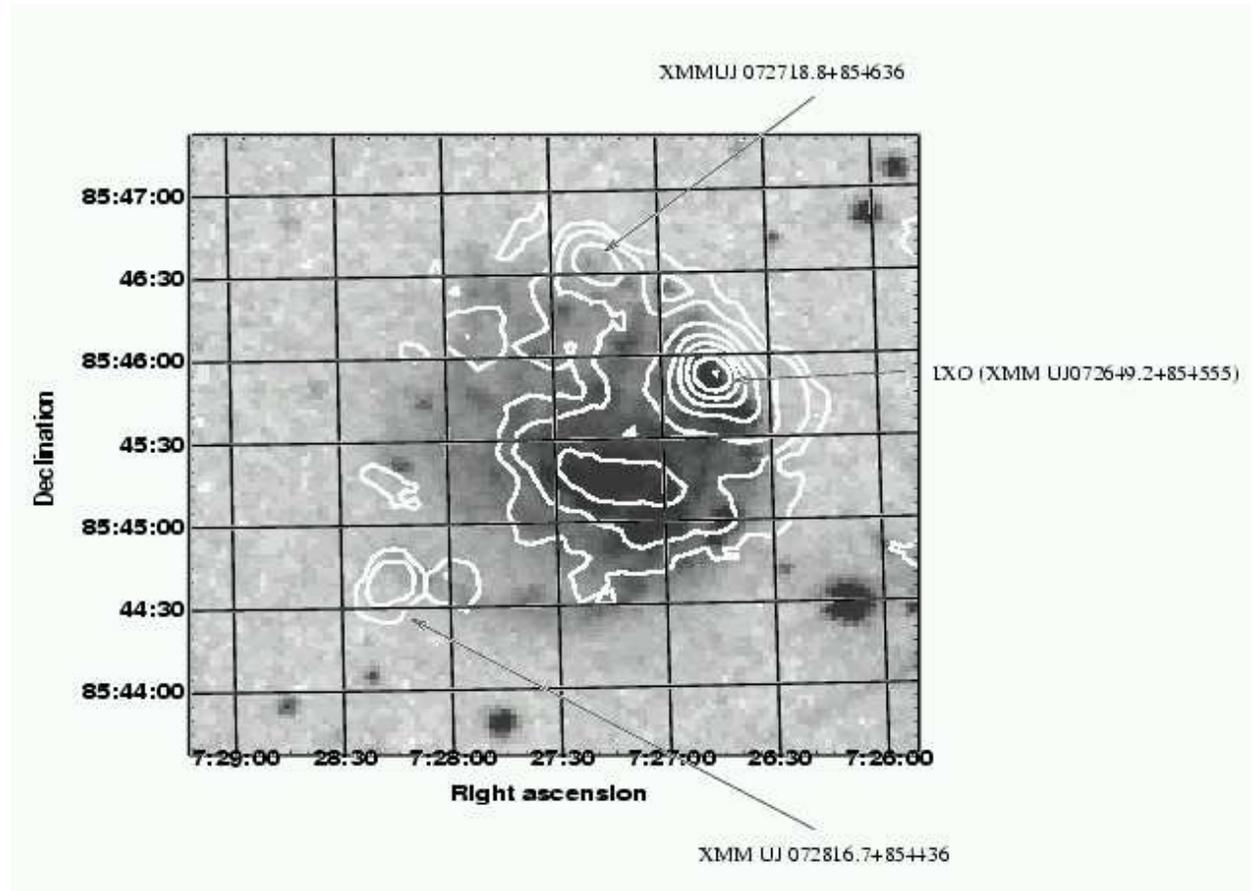}
\caption{The Digital Sky Survey image of NGC~2265 with the XMM X-ray 
contours overlayed.
The XMM data have been smoothed with a $\sigma=15\arcsec$ Gaussian.
}
\end{figure}

\begin{figure}
\plotone{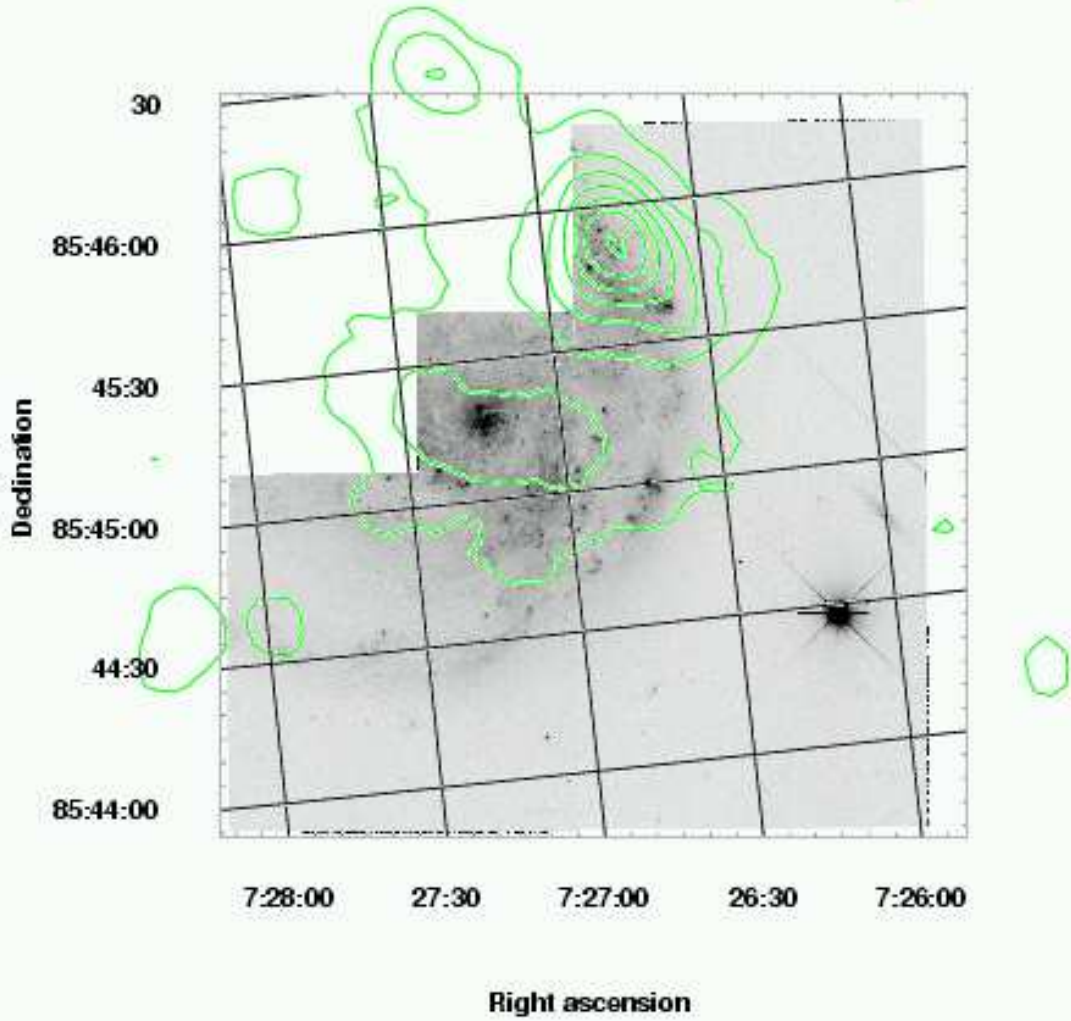}
\caption{The HST image of NGC~2265 with the XMM X-ray contours overlayed.
The XMM data have been smoothed with a Gaussian using a $\sigma=15\arcsec$.
}

\end{figure}

\begin{figure}
\plotone{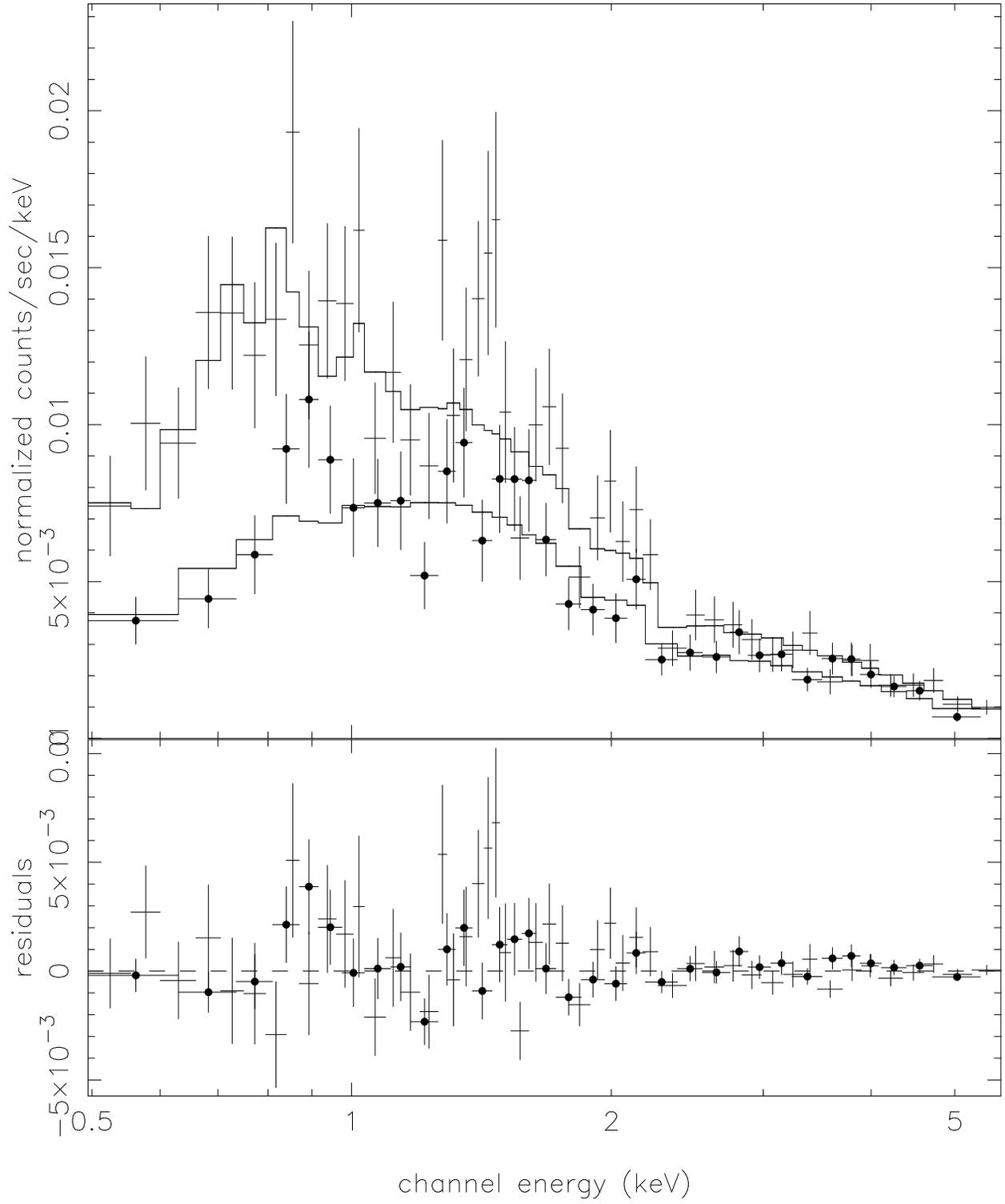}
\caption{The extracted XMM MOS1 and MOS2 (solid points) and the fitted
{\sc XSPEC} diskbb model. }

\end{figure}

\begin{figure}
\epsscale{0.8}
\plotone{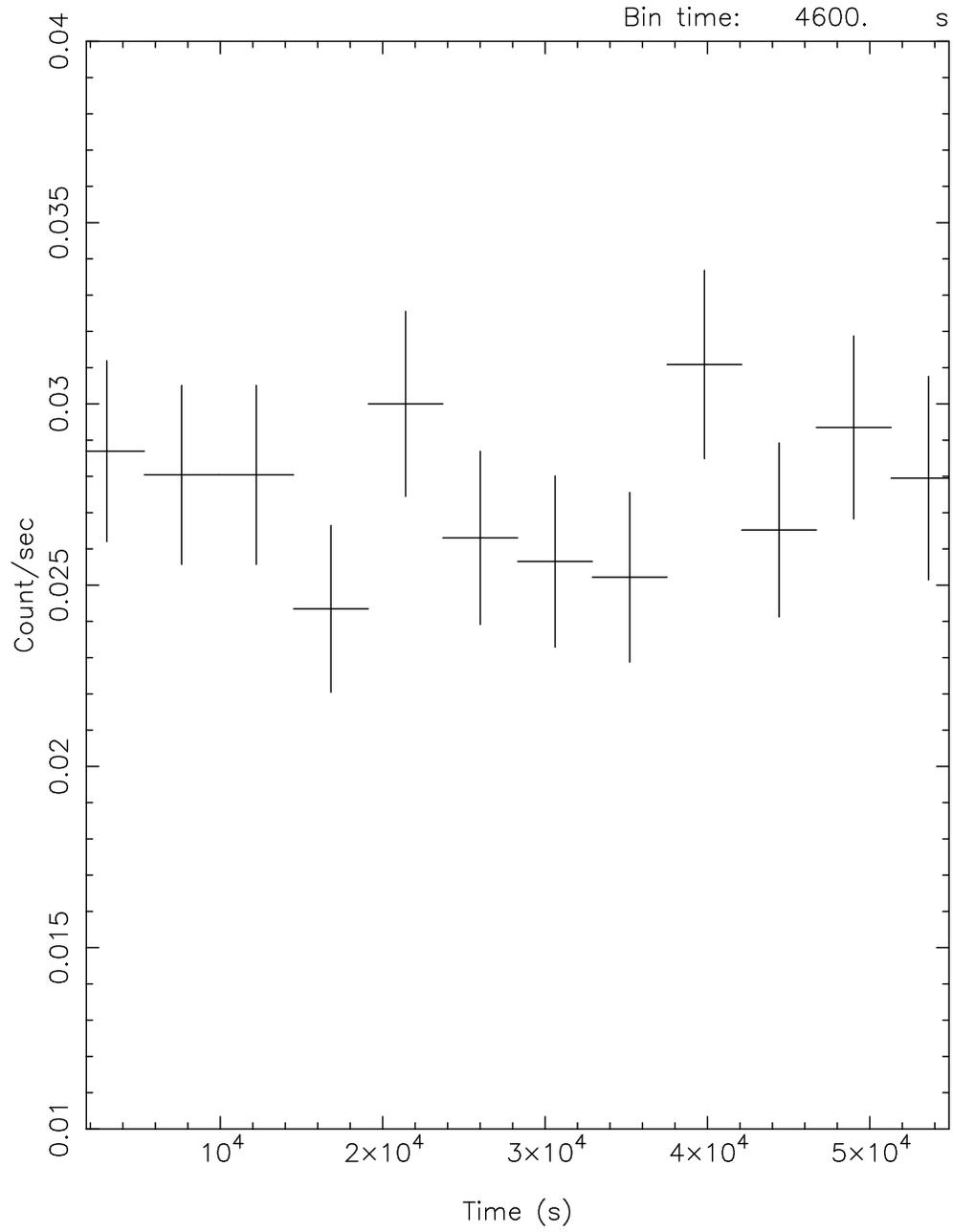}
\caption{The extracted XMM MOS2 lightcurve for the IXO
with 4600 second bins. }

\end{figure}

\begin{figure}
\epsscale{0.8}
\plotone{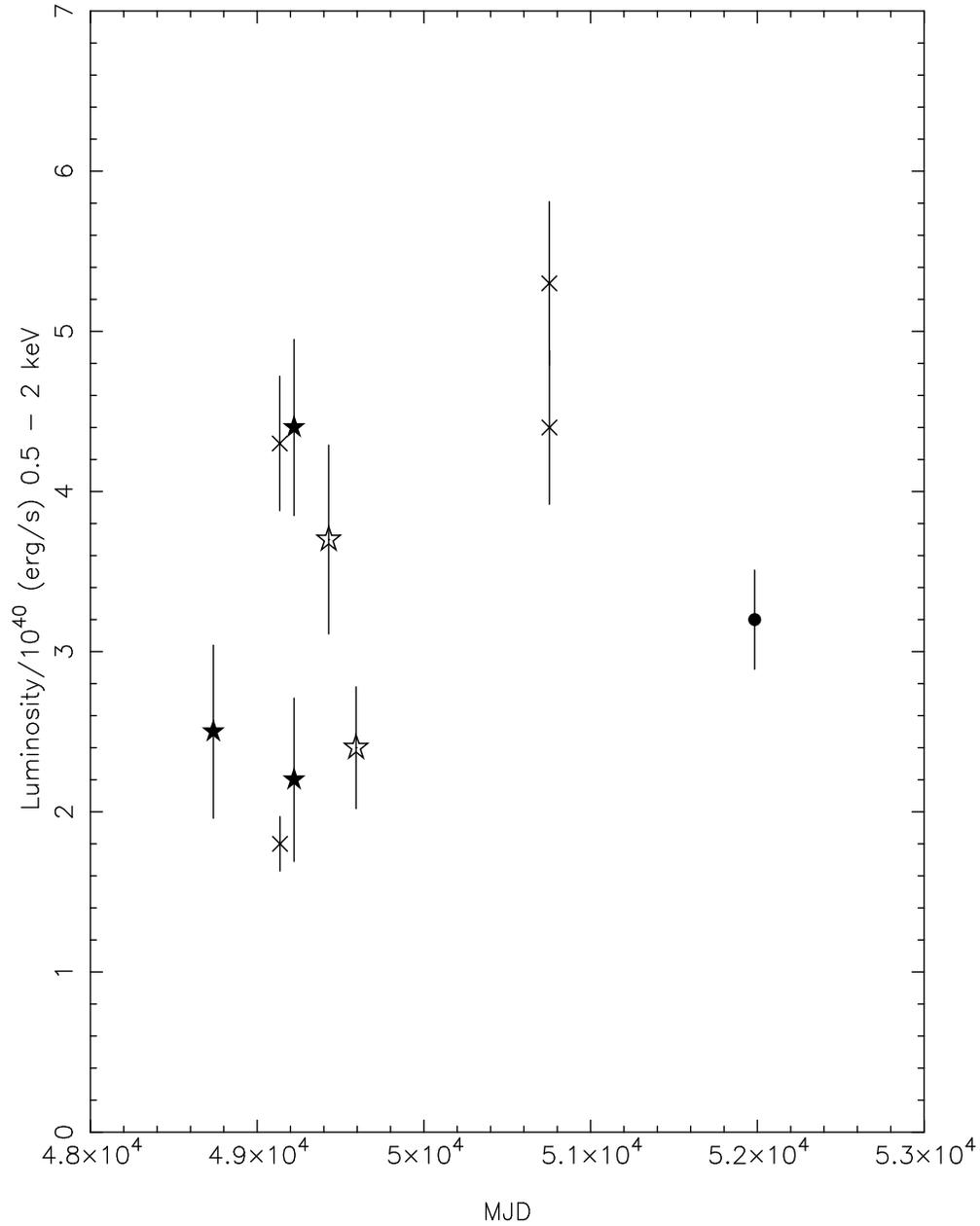}
\caption{The long term lightcurve for the IXO in NGC~2276. The solid stars are
the $ROSAT$ PSPC data, the open stars are the $ROSAT$ HRI data. The $ASCA$ GIS
data are denoted by the crosses and the XMM luminosity is shown with the
solid circle. The flux is in units of 10$^{40}$ erg s${-1}$ in the 0.5 -- 2.0
keV band. The plotted error bars are 1 $\sigma$. It should be noted 
that while the
ASCA data does not isolate the IXO we assume that the preponderance 
of the X-ray
emission is from the IXO. The upper limit for the
Einstein IPC data is not shown here for clarity. }

\end{figure}

\end{document}